# Predictability of Irregular Human Mobility

Travel Patterns of International and Domestic Visitors

Kewei Zhang, Minkyoung Kim, Raja Jurdak, Dean Paini

Commonwealth Scientific and Industrial Research Organisation (CSIRO), Australia

{Kewei.Zhang, Minkyoung.Kim, Raja.Jurdak, Dean.Paini}@csiro.au

## ABSTRACT

Understanding human mobility is critical for decision support in areas from urban planning to infectious diseases control. Prior work has focused on tracking daily logs of outdoor mobility without considering relevant context, which contain a mixture of regular and irregular human movement for a range of purposes, and thus diverse effects on the dynamics have been ignored. This study aims to focus on irregular human movement of different meta-populations with various purposes. We propose approaches to estimate the predictability of mobility in different contexts. With our survey data from international and domestic visitors to Australia, we found that the travel patterns of Europeans visiting for holidays are less predictable than those visiting for education, while East Asian visitors show the opposite patterns, *i.e.*, more predictable for holidays than for education. Domestic residents from the most populous Australian states exhibit the most unpredictable patterns, while visitors from less populated states show the highest predictable movement.

## 1 INTRODUCTION

Understanding human mobility is of fundamental significance due to its important implications in many fields from urban planning [9], to traffic forecasting [11], to transportation management [3], and to infectious disease prediction and control [5, 8]. Human mobility studies have been fueled by collecting and tracking a wide spectrum of human mobility patterns with the help of technical advances in social sensing and mobile devices [7], such as geo-tagged microblogs [6], call data records [10], GPS logs [12], and wireless network transmissions [4]. For example, Twitter data has been proposed as a proxy for human mobility, as it captures rich features of dynamics such as intra- and inter-city movement patterns [6], while mobile phone data can be used for estimating the potential predictability of human mobility [10].

Despite recent advances in studying human mobility, most prior work has focused on tracking daily logs of outdoor mobility, without considering the relevant context and thus ignoring diverse effects on the dynamics. For instance, such daily life logs contain a mixture of regular and irregular human movement, and thus making it difficult to define mobility dynamics. Also, populations with different socio-economic background can lead to forming dissimilar nationwide movement trajectories. Finally, human mobility is driven on a wide variety of purposes, and different purpose-driven mobility leads to distinct patterns. Therefore, it is essential to consider such diverse effects on the dynamics for an accurate and deeper understanding of human mobility.

In this respect, we focus on irregular human movement of different meta-populations with various purposes, and we propose approaches to estimate the predictability of irregular mobility in different context by using information-theoretic measures. That is, we jointly consider three major aspects of human mobility: (1) irregularity, (2) meta-population, and (3) purpose. For this study, we investigate survey data from international and domestic visitors to Australia during a period of more than 10 years. Our data consists of visitor information (*e.g.*, country origin or home location for each international and domestic visitor, respectively), travel purpose, trajectory, period, and so forth, which allows us to disaggregate the distinct movement patterns of different meta-populations.

We find that the travel patterns of Europeans visiting for their holidays are less predictable than when they visit for education, while East Asian visitors show the opposite trends, *i.e.*, more predictable for holidays than for education. Among all the states in Australia, domestic residents from the most populous states exhibit the most unpredictable patterns, while visitors from less populated states show the highest predictable movement. That is, different irregular travel patterns between meta-populations (nationwide or worldwide populations) are likely led by dissimilar socio-economic backgrounds: *e.g.*, centrality-periphery of domestic residential areas and different cultural/social/economic systems between countries.

To the best of our knowledge, this is the first study to discover irregular human mobility dynamics with rich context in terms of different meta-populations (statewide domestic residents and continent-wide international visitors) and representative travel purposes (holiday, business, education). We expect that this study is readily applicable to analyze travelers' movement patterns in different countries and can be an initial step towards identifying diverse socio-economic factors that govern human mobility dynamics.

## 2 DATA AND METHODS

In this section, we explain our data collection, basic statistics of the collected datasets, and finally introduce our methods for estimating the predictability of irregular human mobility.

### 2.1 Data Description

**International Visitor Data.** The International Visitor Survey (IVS) is conducted by Computer Assisted Personal Interviewing (CAPI) in the departure lounges of the eight major international airports in Australia: Sydney, Melbourne, Brisbane, Cairns, Perth, Adelaide, Darwin and the Gold Coast [1]. The collection period of IVS data spans 11 years from 2005 to 2015, sampling approximately 40,000 departing international travelers each year, who are aged 15 years and over. This data consists of traveler information (*e.g.*, nationality, gender, age group), the time period of travel, departing flight information, and more importantly, a series of locations where they have been to during their stay in Australia with associated reasons and activities in a SA2 [1] level. Detailed data statistics is presented in Table 1.

Table 1: Data description of the International Visitor Survey (IVS) and National Visitor Survey (NVS).

| Data | Collection Period | #Visits | #Persons | #Visits per Person | #Days per Person | Holiday (%) | Business (%) | Education (%) | Relatives / Friends (%) |
|---|---|---|---|---|---|---|---|---|---|
| IVS | 2005 - 2015 | 1,274,903 | 442,445 | 2.88 | 48.92 | 50.6 | 7.6 | 8.4 | 21.0 |
| NVS | 1998 - 2015 | 751,184 | 588,323 | 1.28 | 4.22 | 33.2 | 16.0 | 0.7 | 32.6 |

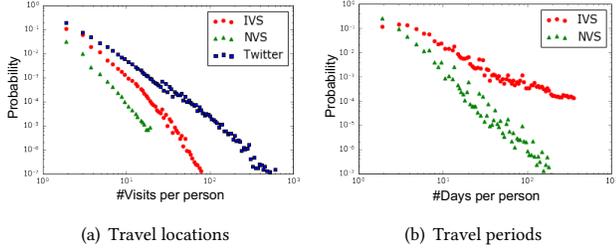

(a) Travel locations  (b) Travel periods

Figure 1: Distributions of each individual's (a) visits and (b) travel days from the IVS and NVS datasets.

**Domestic Visitor Data.** The National Visitor Survey (NVS) captures traveling information of Australian residents from 1998 to 2015 [2]. NVS sample sizes are approximately 33,000 domestic travelers each year, which are comparable to IVS. This data includes trip purpose, travel party type, the number of nights at stopovers, home and stop over locations in a SA2 level, and the date returned from the trip. In contrast to IVS data, NVS data has a longer time period of collection and consists of more detailed travel information. This dataset makes up the basis of analyzing domestic visitor travel patterns. Table 1 describes the datasets in more detail.

### 2.2 Fundamental Statistics

Two fundamental aspects of human movement, trajectory and period, are investigated with our IVS and NVS datasets in terms of the number of visiting locations and travel days.

Accordingly, Figure 1(a) shows the distributions of each individual's visiting locations, separately for international (red circle) and domestic tourists (green triangle), and Twitter users (blue square). Here, Twitter data contains 79,271 accounts with precise geo-locations over the whole year of 2015, which is presented only for the comparison with our datasets. As the figure shows, they are all heavy tailed, indicating that human movement likely covers only a few locations but exceptionally with a large number of locations explored by a few individuals. In comparison, the domestic tourists come up with the shortest tail (maximum 13 visit entries), while international tourists and Twitter users exhibit longer tails in that order (up to 63 and 398 visit entries, respectively). On the other hand, Figure 1(b) shows the heavy tailed travel period distributions between international and domestic visitors. The longer tail distribution of IVS indicates that international visitors likely stay longer than domestic travelers, which is also shown in Table 1 where the average visit time span of international visitors is ten times more than that of domestic travelers. Overall, international tourists visit more locations for a longer time span than domestic tourists.

In terms of travel purposes, holiday is the major reason for international visitors as shown in Figure 2(a), whose proportion in particular is dominant in North Territory (NT), Tasmania (TAS), and Queensland (QLD). On the other hand, two major purposes (holiday and visiting relatives or friends) of domestic tourists lead to almost the same proportions of visitors across the states as shown in Figure 2(b). That is, international tourists likely visit popular areas due in part to limited time and opportunities, leading to wider popularity disparity. On the contrary, domestic tourists pay attention to nationwide areas due to lifetime travel opportunities, leading to balanced popularity across the states. Particularly, domestic residents visit to NT and Australian Capital Territory (ACT) mainly for business, which is largely different from international visitors to these states.

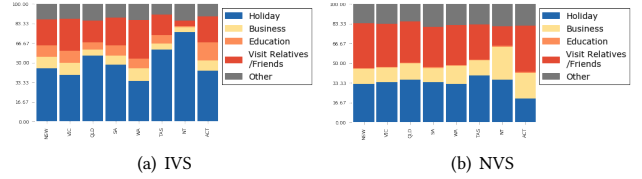

(a) IVS  (b) NVS

Figure 2: Statewide visitor distributions of four major travel purposes (Holiday, Business, Education, Visit Relatives or Friends) for our two datasets, (a) IVS and (b) NVS.

### 2.3 Estimation of Predictability

In this section, we explain how to estimate the predictability of irregular human movement by using three different types of entropy measures, which are summarized below.

(1) **Information of Visits:**

$$S_i^{\text{rand}} \equiv \log_2 N_i \ , \tag{1}$$

where $N_i$ is the number of distinct locations visited by a meta-population $i$ (e.g., state, country, world region), capturing the information about the target population's whereabouts (randomness of visiting locations).

(2) **Uncertainty of Visits:**

$$S_i^{\text{unc}} \equiv -\sum_{j=1}^{N_i} p_i(j) \log_2 p_i(j) \ , \tag{2}$$

where $p_i(j)$ is the probability that location $j$ was visited by a meta-population $i$, characterizing the uncertainty of visiting patterns.

(3) **Uncertainty of Displacement Distance:**

$$S_i^{\text{weight}} \equiv -\sum_{e=1}^{N_i} \phi(e) p_i(e) \log_2 p_i(e) \ , \tag{3}$$

---
[1]Statistical area level 2 (SA2) is one of the spatial units defined under the Australian Statistical Geography Standard (ASGS).

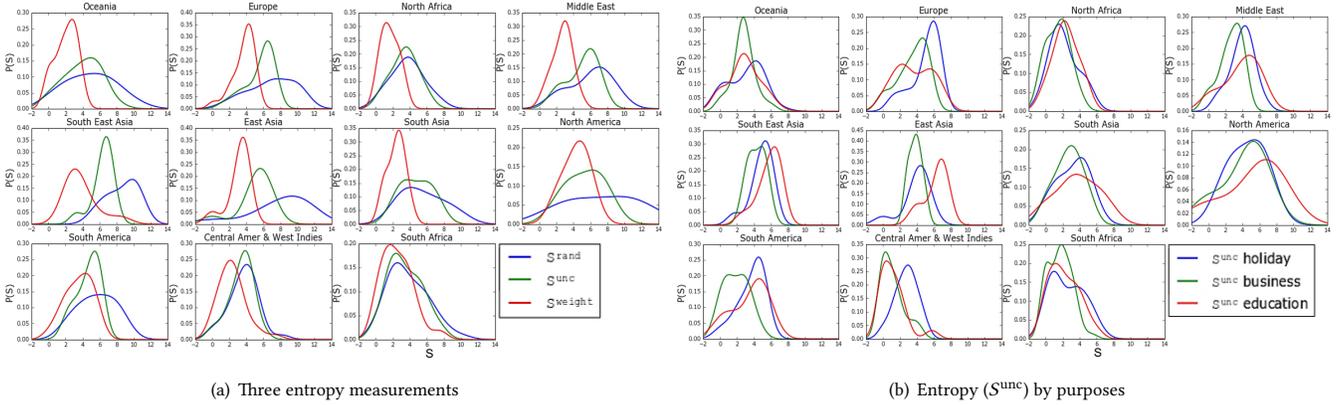

(a) Three entropy measurements  (b) Entropy ($S^{unc}$) by purposes

Figure 3: (a) The distributions of estimations by three measurements such as $S_i^{rand}$ (blue) in Eq. (1), $S_i^{unc}$ (green) in Eq. (2), and $S_i^{weight}$ (red) in Eq. (3) for all international visitors from 11 regions. (b) The distributions of estimations by the entropy $S_i^{unc}$, separated by international visitors' travel purposes: holiday (blue), business (green), education (red).

where $p_i(e)$ is the probability that a route $e$ (i.e., an edge connecting two distinct locations) was traveled by a meta-population $i$, and $\phi(e) = \frac{d(e)}{D_{guard}}$ is a weight function. Here, $d(e)$ is the distance of a route $e$, and $D_{guard}$ is a normalization factor varying with geographical characteristics.

This measure calculates the weighted entropy by weighting each event's outcome over distance and thus distinguishing the probability distributions of meta-populations' traveling patterns but with different distances. This is not allowed in the entropy in Eq. (2), since it is outcome independent.

## 3 UNCERTAINTY OF IRREGULAR HUMAN MOBILITY

As discussed in the introduction section, we consider the context of irregular human mobility in terms of meta population and travel purposes, which helps better understand mobility dynamics.

### 3.1 International Visitor Movement

We first categorize the home countries of international visitors into 11 region groups based on the geographical locations of all countries for high level interpretations of mobility patterns. In order to characterize the inherent predictability of visitor movement from each region, we estimate three different types of entropy $S_i^{rand}$, $S_i^{unc}$, and $S_i^{weight}$ for each group $i$, and their probability distributions are presented in Figure 3(a). Evidently, the order of estimated entropy across 11 world regions is $S_i^{rand} > S_i^{unc} > S_i^{weight}$ when $D_{guard} = 1000km$ in Eq. (3). Here, we set $D_{guard}$ as 1000 $km$ which represents the estimated average distance between all metropolitan cities in Australia.

**General Predictability.** Interestingly, the top two regions (Europe and East Asia) stand out compared to the others, exhibiting greater heterogeneity ($S_i^{rand}$) of visiting locations due in part to larger visitor sizes than the others. More specifically, the mode value of $S_i^{rand}$ in East Asia is larger than the one in Europe, while the modes of $S_i^{unc}$ and $S_i^{weight}$ in East Asia are all smaller than those in Europe.

These imply that East Asian visitors show higher heterogeneity (i.e., lower predictability) in the number of visited locations but lower uncertainty (i.e., higher predictability) regarding the distributions of their travel locations and distances than European visitors. The visiting patterns ($S_i^{unc}$) of South and North African tourists are the most predictable (followed by Oceania and South America), while those of European, North American, and South East Asian tourists are the most unpredictable. Regarding the uncertainty of displacement distance ($S_i^{weight}$), tourists coming from Central Amer & West Indies and Africa (North and South) show a tendency to travel to more certain locations with shorter distances, in contrast to North American tourists.

**Predictability with Different Travel Purposes.** However, these general predictabilities can vary with travel purposes. Accordingly, Figure 3(b) depicts the entropy ($S_i^{unc}$) distributions for each region, separated by the three representative travel purposes: holiday (blue), business (green), and education (red). In general, business travel is the most predictable across all world regions. This can be explained by the geographic concentration of businesses in Australia, such as Sydney as a financial center and Perth as an oil, petrochemical and mining business district.

We also identify that education is a dominant purpose of tourists coming from Middle East, South East Asia, East Asia, and South America, which explains relatively higher entropy of their educational visits than the others. Interestingly, the top two regions, Europe and East Asia exhibit opposite trends in predictabilities of holiday and education mobility patterns. For instance, the travel patterns of Europeans visiting for their holidays are less predictable (higher entropy) than those for education, while East Asian visitors are more predictable (lower entropy) for holidays than for education.

### 3.2 Domestic Visitor Movement

We now estimate the predictability of domestic visitor movement, as conducted in the previous section. We also categorize the home locations of domestic residents into eight states in Australia for high level interpretations of their mobility patterns: New South

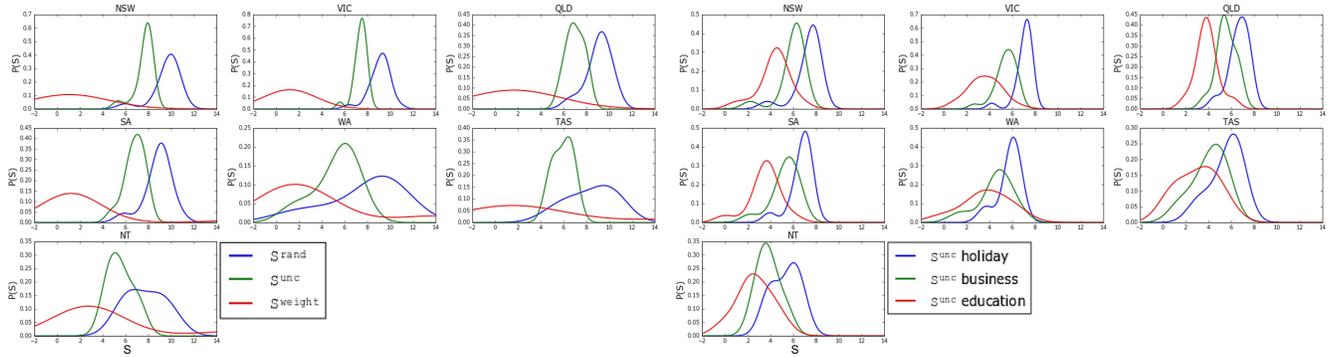

(a) Three entropy measures      (b) Entropy ($S^{\text{unc}}$) by purposes

Figure 4: (a) The distributions of estimations by three measurements such as $S_i^{\text{rand}}$ (blue) in Eq. (1), $S_i^{\text{unc}}$ (green) in Eq. (2), and $S_i^{\text{weight}}$ (red) in Eq. (3) for domestic visitors from seven states in Australia: New South Wales (NSW), Victoria (VIC), Queensland (QLD), South Australia (SA), West Australia (WA), Tasmania (TAS), and Northern Territory (NT). (b) The distributions of estimated entropy $S_i^{\text{unc}}$, separated by domestic visitors' travel purposes: holiday (blue), business (green), education (red).

Wales (NSW), Victoria (VIC), Queensland (QLD), South Australia (SA), West Australia (WA), Tasmania (TAS), Northern Territory (NT), and Australian Capital Territory (ACT).

Figure 4(a) accordingly presents the distributions of three different types of estimated entropy $S_i^{\text{rand}}$, $S_i^{\text{unc}}$, and $S_i^{\text{weight}}$ with $D_{\text{guard}} = 3000km$. Note that, $D_{\text{guard}} = 3000km$ represents the estimated average distance between capital cities of each state in Australia, as domestic tourists prefer longer distance travel (interstate movement) than international visitors.

**General Predictability.** As in the case of international visitor movement, the order of estimated entropy across the states is $S_i^{\text{rand}} > S_i^{\text{unc}} > S_i^{\text{weight}}$, but the weight entropy $S_i^{\text{weight}}$ is more widely distributed than one for international visitors. That is, the likelihood of travel distance has higher variance. Interestingly, domestic residents from the top populous states such as NSW, VIC, and QLD exhibit the most unpredictable patterns in terms of both visiting locations ($S_i^{\text{rand}}$) and their distributions ($S_i^{\text{unc}}$), while visitors from less populated states such as NT, TAS, SA, and WA show the highest predictable movement. That is, irregular travel patterns of domestic residents can be influenced by socio-economic factors. For instance, domestic visitors whose residential areas are closer to the population concentration likely have more options to choose ranges of travel routes than ones from less populated areas due in part to relatively well developed transportation infrastructures.

**Predictability with Different Travel Purposes.** When considering travel purposes, domestic visitors show generally similar patterns across the states such that $S^{\text{holiday}} > S^{\text{business}} > S^{\text{education}}$ ($S^{\text{holiday}} \equiv S_{i,\text{holiday}}^{\text{unc}}$ for brevity). That is, visitors for education are the most predictable than those for business and holiday purposes in that order. Interestingly, the mode values of entropy distributions for education are all around 4 except for NT due to the limited number of universities in NT, while the ones for business and holidays vary across the states. Such similar uncertainty of mobility patterns for education reflects balanced educational opportunities across the states in Australia.

## 4 CONCLUSION

We proposed approaches to estimate the predictability of irregular mobility by using information-theoretic measures and by considering rich context in terms of meta-populations with different socio-economic backgrounds and various travel purposes. This enabled us to capture the interplays among mobility, population, and purpose. This study sheds a light on the consideration of diverse socio-economic effects on human mobility dynamics such as centrality-periphery of domestic residential areas and different cultural/social/economic systems between countries.

One direction of future work is to identify and incorporate major features of regular and irregular human movement to a new model for predicting mobility patterns.